\documentclass[aps,pra,reprint,superscriptaddress,floatfix,showpacs,showkeys]{revtex4-2}
\usepackage{xcolor}
\usepackage{graphicx}
\usepackage{dcolumn}
\usepackage{siunitx}
\usepackage{float}
\usepackage{amsmath}
\usepackage{bm}
\begin{document}
\title{Protected Ion Beam Fabrication of Two-Dimensional Transition Metal Dichalcogenides based Photonic Devices}

\author{Lekshmi Eswaramoorthy}
 \affiliation{Laboratory of Optics of Quantum Materials, Department of Physics, Indian Institute of Technology Bombay, Powai, Mumbai -- 400076, India}
 \affiliation{Department of Materials Science and Engineering, Monash University, Clayton, Victoria 3800, Australia}
 \affiliation{IITB-Monash Research Academy, Indian Institute of Technology Bombay, Powai, Mumbai -- 400076, India}
\author{Parul Sharma}
\author{Brijesh Kumar}
\author{Abhay Anand}
\author{Anuj Kumar Singh}

 \affiliation{Laboratory of Optics of Quantum Materials, Department of Physics, Indian Institute of Technology Bombay, Powai, Mumbai -- 400076, India}

\author{Sudha Mokkapati}
 \affiliation{Department of Materials Science and Engineering, Monash University, Clayton, Victoria 3800, Australia}
 
\author{Anshuman Kumar}
\email{anshuman.kumar@iitb.ac.in}
 \affiliation{Laboratory of Optics of Quantum Materials, Department of Physics, Indian Institute of Technology Bombay, Powai, Mumbai -- 400076, India}
\date{\today}

\begin{abstract}
Two-dimensional (2D) transition metal dichalcogenides are pivotal for next-generation photonic devices due to their exceptional optical properties and strong light-matter interactions. However, their atomic thinness renders them susceptible to damage during nanoscale fabrication. Focused ion beam technology, while offering precise defect engineering for tailoring optoelectronic properties, often induces collateral damage far beyond the target region, compromising device performance. This study addresses the critical challenge of preserving the intrinsic optical characteristics of 2D TMDCs during FIB patterning. We demonstrate that conventional dielectric encapsulation fails to protect 2D TMDCs from gallium ion-induced damage, leading to persistent defects and quenched optical responses in patterned microstructures. In contrast, polymeric encapsulation with PMMA (polymethyl methacrylate) effectively mitigates damage by acting as a sacrificial layer that absorbs ion impact, thereby preserving the optical properties of the underlying TMDC. Furthermore, we leverage XeF$_2$-assisted Ga ion beam direct patterning, which significantly reduces collateral damage, minimizes Ga ion implantation, and enables precise anisotropic material removal, yielding ultra-smooth sidewalls critical for high-quality photonic resonators. This combined approach of PMMA encapsulation and XeF$_2$-assisted FIB patterning offers a robust, cost-effective, and scalable single-step fabrication route for integrating 2D TMDCs into high-performance photonic devices, thereby maintaining their intrinsic optical functionality essential for advancing quantum technologies and compact optical circuits.

\end{abstract}

\maketitle

\section{\label{sec:level1}Introduction}

Two-dimensional (2D) transition metal dichalcogenides have garnered significant attention due to their exceptional optical properties and strong light-matter interactions, positioning them as promising candidates for next-generation integrated photonic devices. Their tunable bandgap and anisotropic electrical and optical properties enable diverse applications, including photodetectors, flexible nanoelectronics, and low-loss nanophotonics. A key characteristic is their transition from an indirect to a direct bandgap when thinned to a monolayer, significantly enhancing photoluminescence intensity.

However, the inherent vulnerability of these atomically thin materials to extrinsic effects poses a significant challenge for stable bandgap engineering, necessitating robust strategies to preserve their functionalities during device integration. Focused ion beam technology offers a powerful tool for precise nanoscale fabrication and defect engineering in functional nanomaterials such as 2D TMDCs, enabling the tailoring of desirable optoelectronic properties \cite{ziqi_li_5ab9729b}, \cite{fahrettin_sarcan_fddabd15}, \cite{heng_xu_7cb16888}. The controlled introduction of defect sites via FIB allows for localized tuning of sub-gap absorption, leading to the engineering of quantum emitters and precise tailoring of optical responses at the nanoscale \cite{vighter_iberi_71ffbd48}. This defect engineering approach provides a scalable method to controllably introduce defects, thereby modulating exciton dynamics and fine-tuning the optical properties of 2D TMDCs \cite{michael_g__stanford_86b43c7a}, \cite{r__wilhelm_2323d1cb}. Direct patterning of 2D materials is crucial for seamless integration into photonic circuits, enabling intricate device architectures and localized modification of optical responses, while circumventing issues like mechanical damage, contamination, and defects associated with traditional layer transfer processes \cite{marcel_weinhold_029080e2}.

Despite the advantages of FIB for nanoscale patterning, the damage caused by the FIB irradiation and milling process to these delicate, atomically thin materials, especially in extended areas beyond the FIB target, has not yet been fully characterized \cite{fahrettin_sarcan_fddabd15}. Understanding the correlation between these lateral ion beam effects and the optical properties of 2D TMDCs is crucial for designing and fabricating high-performance optoelectronic devices \cite{fahrettin_sarcan_fddabd15}. Previous studies have investigated the impact of heavy ions and gallium ion beams on the optical properties of monolayer WS$_2$, identifying distinct zones of damage and alterations in photoluminescence and Raman spectra \cite{fahrettin_sarcan_fddabd15}. Similar observations have been made regarding defect introduction in MoS$_2$ and WSe$_2$ using helium and gallium ion beams, showing modifications to electrical and optical characteristics \cite{zahra_fekri_d4e92c29}, \cite{heng_xu_7cb16888}.

\begin{figure}[ht]
    \centering
    \includegraphics[width=\linewidth]{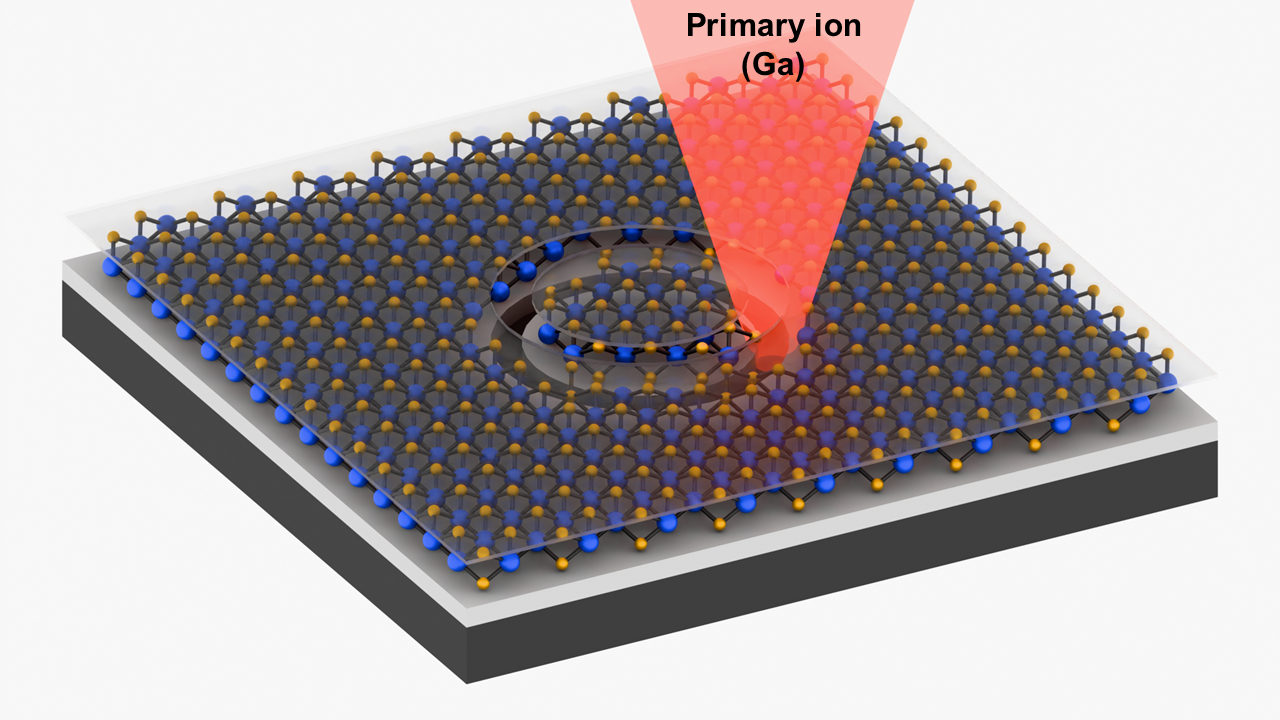}
    \caption{Schematic of Ga-Focused Ion Beam (FIB) patterning on a dielectric encapsulated 2D TMDC layer}
    \label{fig:1}
\end{figure}

In this work, we investigate the lateral damage in large-area monolayer MoS$_2$ caused by the gallium focused ion beam milling process. Our findings reveal that even remote regions, not directly exposed to the primary ion beam, experience significant alterations in their photoluminescence and Raman spectra, indicating a widespread impact on crystal quality and electronic structure \cite{fahrettin_sarcan_fddabd15}. Expanding on these observations, we previously noted that gallium ion irradiation on CVD-grown 2D WSe$_2$ and MoS$_2$ on SiO$_2$/Si substrates resulted in a loss of Raman and PL signatures as far as 100 µm away from the ion bombardment site. Figure \ref{fig:1} provides a schematic of the standard Focused Ion Beam (FIB) patterning process using a primary Gallium (Ga) ion beam on a dielectric encapsulated TMDC layer. While direct sputtering facilitates precise material removal, the high kinetic energy of the ions induces collateral damage via ion channeling, secondary cascades, and implantation that extends significantly beyond the milled region. To address this challenge, our current approach utilizes a combination of polyemric encapsulation and XeF$_2$-assisted gallium ion FIB patterning of the 2D monolayer. This method effectively retains the optical properties of the TMDCs, significantly mitigating the widespread damage typically observed with conventional FIB techniques and preserving the integrity and functionality of the 2D material beyond the patterned region \cite{alex_belianinov_4195812f}, \cite{mobashar_hassan_691d58ff}, \cite{rakesh_d__mahyavanshi_55825eaa}. This preservation is paramount for the development of integrated photonic devices, where maintaining the optical quality of 2D materials after patterning is essential for device performance and stability. Encapsulation strategies, particularly with materials like PMMA and hBN, are critical for mitigating environmental degradation and ensuring the long-term integrity and spectral stability of engineered optical properties, paving the way for robust and efficient integrated photonic devices \cite{mobashar_hassan_691d58ff}, \cite{shanshan_wang_a8f9b1c3}. 

\section{Methodology}

This study systematically investigates the preservation of tailored optical properties in 2D Transition Metal Dichalcogenides following heavy ion beam irradiation for integrated photonic applications. Our methodology encompasses the chemical vapor deposition synthesis of monolayer TMDCs, substrate preparation, various encapsulation strategies, precise focused ion beam patterning, and subsequent optical characterization.

Monolayer MoS$_2$ was synthesized using a custom-built double-zone CVD furnace. A 285 nm thick silicon dioxide (SiO$_2$) layer, deposited via thermal oxidation on a silicon substrate, served as the robust dielectric foundation. This SiO$_2$/Si substrate was chosen to provide a stable, electrically insulated base, minimizing substrate-induced doping effects and ensuring consistent optical properties. However, the thermal expansion coefficient of SiO$_2$ often differs from that of materials like MoS$_2$, which can lead to strain in CVD-grown MoS$_2$ crystals on SiO$_2$ substrates \cite{wenjuan_huang_efac4fcc}. This strain can impact the performance and properties of the materials \cite{kai_liu_6576b20e}.

\begin{figure*}[ht]
    \centering
    \includegraphics[width=1\linewidth]{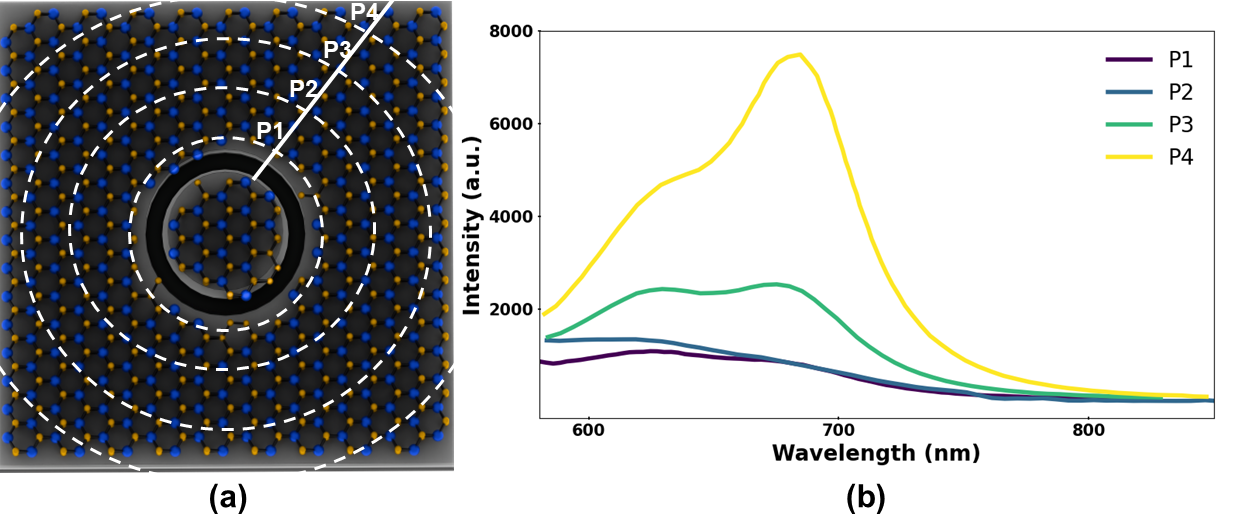}
    \caption{Radial extent of optical damage in $\text{Al}_2\text{O}_3$ encapsulated 2D $\text{MoS}_2$ following Ga-FIB patterning.(a) Schematic showing photoluminescence (PL) measurement points ($\text{P}1$ to $\text{P}4$) at increasing radial distances from the ion-milled region (b) Corresponding PL spectra demonstrating a severely quenched exciton signal ($\text{P}1$, $\text{P}2$) that only recovers to the intrinsic level ($\text{P}4$) far from the patterned region, quantifying the non-local damage.}
    \label{fig:2}
\end{figure*}

Molybdenum oxide (MoO$_3$) and sulfur powders were used as precursors under controlled temperature and carrier gas flow to promote uniform monolayer growth. The as-grown MoS$_2$ monolayers were subjected to rigorous characterization, including atomic force microscopy and photoluminescence spectroscopy, to confirm their quality and uniformity prior to heavy ion irradiation. These growth parameters ensure high-quality monolayers essential for subsequent heavy ion beam patterning and optical characterization \cite{shisheng_li_cc40a6fb}, \cite{yuqing_li_c8db001a}, \cite{kai_liu_6576b20e}, \cite{lei_tang_9afc927a}.

FIB technology was employed for precise nanoscale fabrication and patterning of the TMDC materials. A commercially available 30 kV dual-beam FIB machine (Thermo Fisher Helios 5 UC) was utilized for patterning and milling. In this system, the ion gun is positioned at 38° to the stage, and the sample is tilted 52° towards the ion gun during patterning. The electron gun is normal to the sample. The overall ion dose at each feature was programmatically controlled via beam exposure duration and beam current.

Direct nanofabrication using FIB can introduce damage, particularly from Ga ion implantation, which alters the optical properties of 2D TMDC materials. The extent of this variation is influenced by beam current and distance from the ion bombardment site \cite{fahrettin_sarcan_fddabd15}. The implanted area typically comprises a severely damaged central region from the primary beam and a less severely damaged outer ring originating from beam tails. Ion irradiation leads to various types of defects: Type I and II (end-of-range defects) are associated with interstitial atoms forming dislocation loops or clusters, while Type III and IV defects are linked to solid-phase epitaxial regrowth in amorphized regions. Type V defects and precipitates form when implanted species exceed their solubility limit. To minimize Ga ion implantation and surface damage, especially in the SiO$_2$ layer, beam currents were restricted to lower values, ensuring more accurate material removal and control over the polishing procedure. Optimizing the incident ion angle and energy can further refine nanostructuring, enabling selective modification of optical properties with minimal collateral damage.

\subsection{\label{sec:level3}Dielectric Encapsulation}
\begin{figure}[h]
    \centering
    \includegraphics[width=\linewidth]{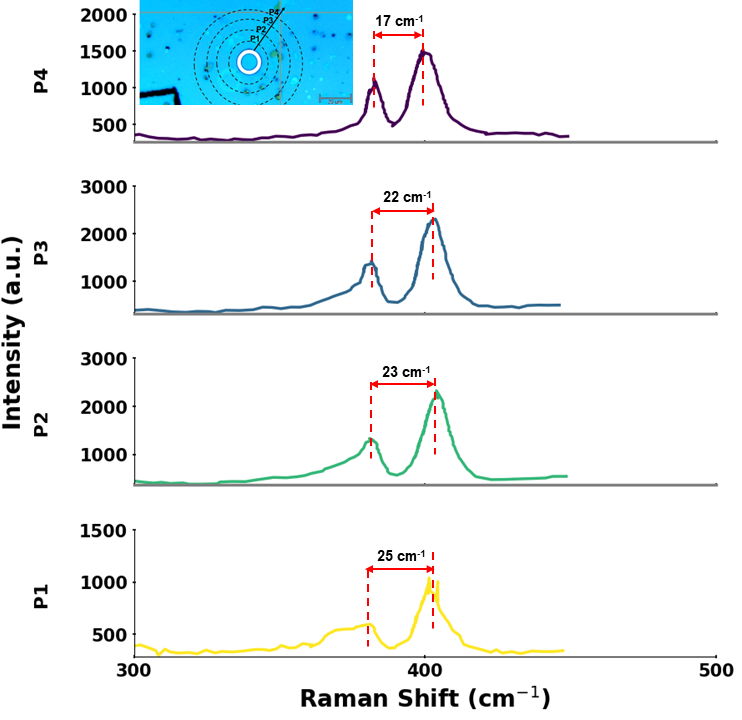}
    \caption{Effect of dielectric encapsulation: Raman spectra of $\text{Al}_2\text{O}_3$ encapsulated $\text{MoS}_2$ show a significant peak shift and broadening closer to the site of ion bombardment. The inset shows the optical micrograph detailing the patterned area and the specific measurement points ($\text{P}1$ to $\text{P}4$).}
    \label{fig:3}
\end{figure}
To mitigate degradation and preserve the optical properties of the 2D TMDCs during and after ion beam processing, various encapsulation layers were explored. Both dielectric (Al$_2$O$_3$ and SiO$_2$) and polymeric thin films were utilized. 20~nm thick dielectric films were grown using Atomic Layer Deposition, while PMMA films were directly spin-coated onto the 2D TMDCs.

\begin{figure*}[htb]
    \centering
    \includegraphics[width=0.4\linewidth]{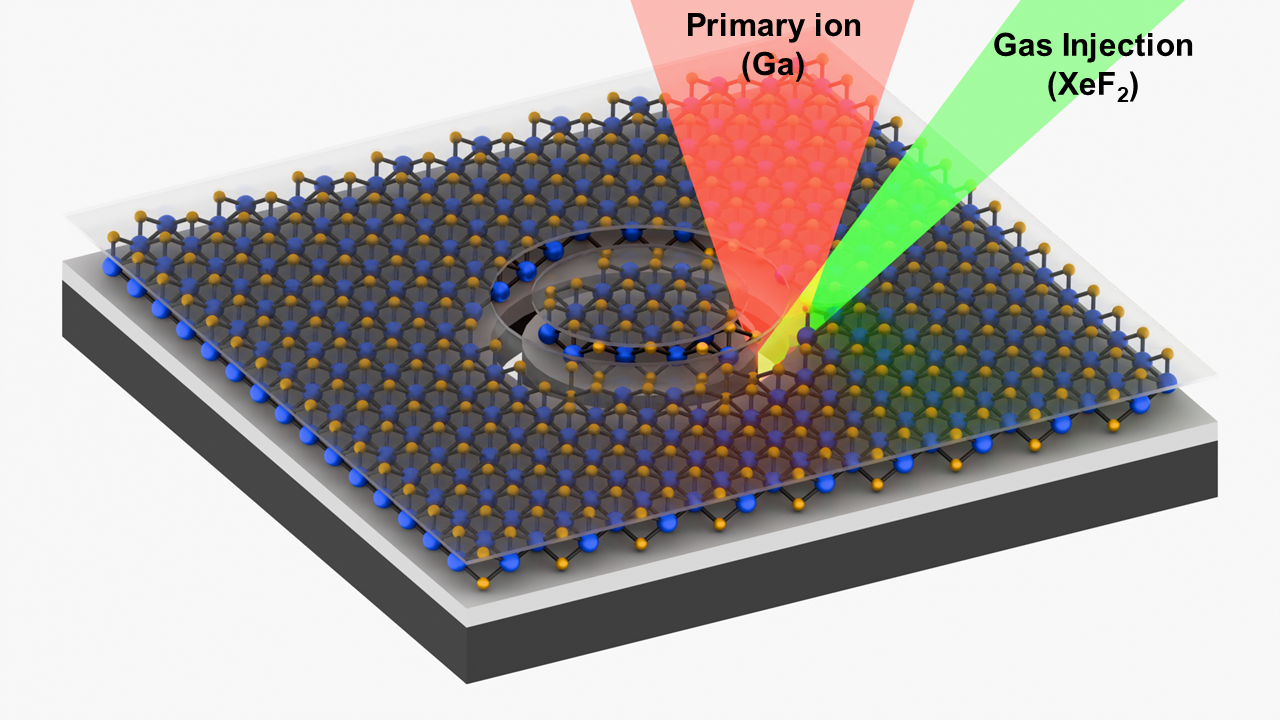}
     \includegraphics[width=0.5\linewidth]{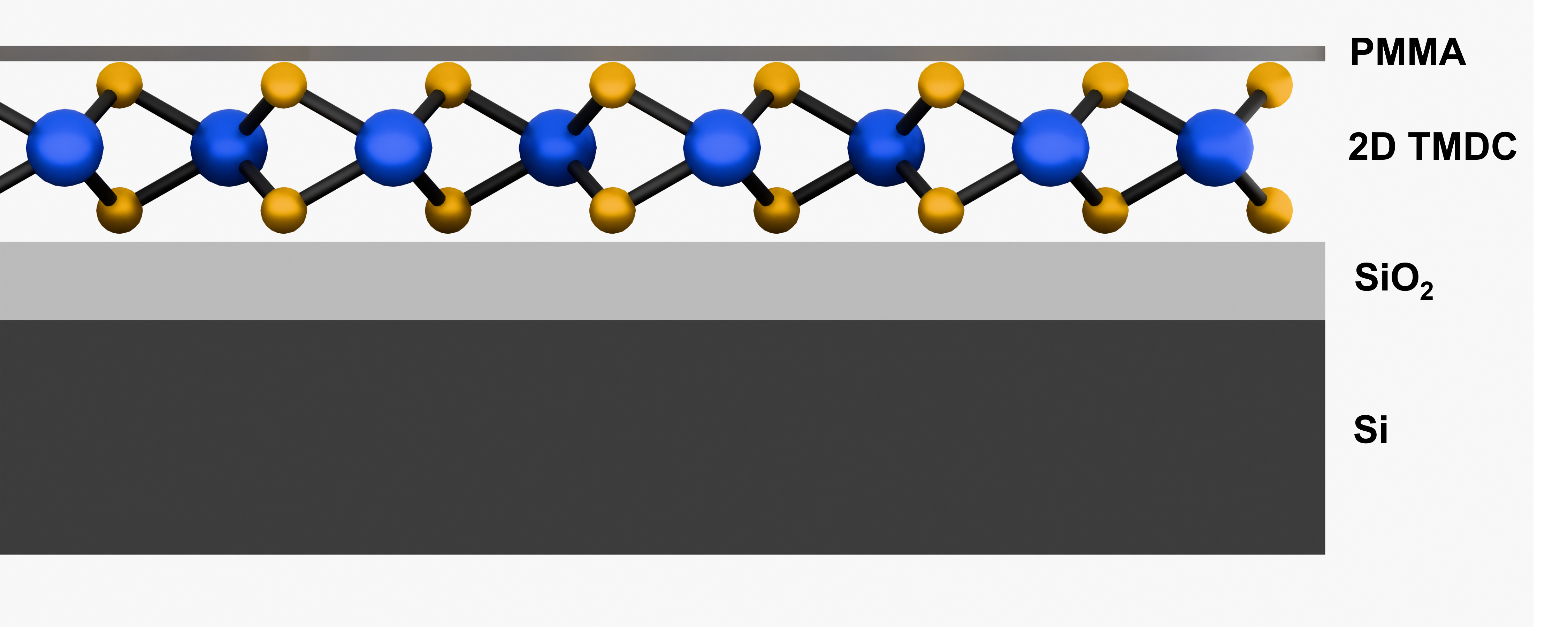}
    \caption{PMMA encapsulation and $\text{XeF}_2$-assisted FIB patterning: Illustration of Gas Injection System (GIS)-assisted FIB patterning using $\text{XeF}_2$, which enhances chemical etching to minimize Ga-ion implantation, yielding ultra-smooth sidewalls. Schematic cross-section shows the protective PMMA layer over the 2D TMDC, acting as a sacrificial layer to absorb ion energy and prevent non-local damage}
    \label{fig:4}
\end{figure*}
Initial investigations revealed that dielectric capping altered the optical properties of the 2D TMDCs, attributed to a doping effect on the monolayer. This doping can significantly influence the electronic band structure and exciton dynamics, modulating photoluminescence quantum yield and optical absorbance. In contrast, polymeric encapsulation with PMMA proved effective in retaining the optical properties of the 2D materials. This protective effect is attributed to PMMA's ability to act as a sacrificial layer, absorbing the impact of gallium ions and minimizing direct damage to the underlying TMDC. The pliable nature of PMMA allows for energy dissipation, acting as a kinetic buffer that reduces momentum transfer to the 2D material layers. Furthermore, PMMA's molecular structure, with its high hydrogen content, can increase inelastic scattering of incident ions, further reducing their energy before reaching the TMDC \cite{rakesh_d__mahyavanshi_55825eaa}, \cite{jothi_priyanka_thiruraman_f3c30039}. This differential protection mechanism underscores the importance of selecting appropriate encapsulation materials based on the specific processing techniques and desired outcomes for integrated photonic devices \cite{wenjuan_huang_efac4fcc}.

\subsection{\label{sec:level2}Polymeric Encapsulation}

To overcome the limitations of conventional FIB patterning and mitigate widespread damage, a combination of PMMA encapsulation and XeF$_2$-assisted Ga-ion FIB patterning was employed to fabricate microdisks. This method reduced the collateral damage and Ga-ion implantation, common issues with direct FIB milling \cite{kyle_mahady_7cf2b611}. The chemical etching in XeF$_2$-assisted patterning enabled precise anisotropic material removal, allowing for the definition of intricate microdisk geometries crucial for resonant optical modes \cite{jangyup_son_b419e2dd}, \cite{fernando_j__urbanos_17a5dca2}. This approach simplifies fabrication into a single step process, minimizes re-deposition of sputtered material, and facilitates the creation of ultra-smooth sidewalls, critical for minimizing scattering losses and enhancing the Q-factor of fabricated photonic resonators \cite{jian_zhou_f55c1f27}.

The optical properties of the TMDC monolayers were characterized primarily using photoluminescence and Raman spectroscopy. These techniques were applied to both as-grown and FIB-patterned samples, with and without encapsulation, to assess the impact of the fabrication processes on crystal quality, electronic structure, and defect formation. PL and Raman spectroscopy are crucial for identifying distinct damage zones, quantifying changes in emission wavelengths, decay lifetimes, and monitoring the effectiveness of encapsulation strategies in preserving the intrinsic optical response of the 2D materials \cite{zahra_fekri_d4e92c29}, \cite{fahrettin_sarcan_fddabd15}. This includes monitoring the loss or retention of PL and Raman signatures in regions remote from the direct ion bombardment site.

\section{Results and inferences}

Our experimental investigations focused on assessing the efficacy of different encapsulation strategies and patterning techniques in preserving the optical properties of 2D TMDCs during focused ion beam processing for integrated photonic device fabrication.

When the encapsulated monolayer was subjected to FIB patterning for microdisk fabrication, we observed that dielectric capping (Al$_2$O$_3$ and SiO$_2$) failed to adequately protect the 2D TMDC from the detrimental influence of gallium ions on its optical properties. To quantify this non-local damage, Photoluminescence (PL) measurements were conducted on patterned $\text{MoS}_2$. As shown in Figure~\ref{fig:2}, the exciton emission signal was measured at four points ($\text{P}1$ to $\text{P}4$) spaced at $\mathbf{10~\mu\text{m}}$ intervals, with $\text{P}1$ beginning $10~\mu\text{m}$ away from the edge of the ion-milled region. The signal is severely quenched near the patterned center ($\text{P}1$ and $\text{P}2$), recovering to its intrinsic level only at large radial distances ($\text{P}4$), which is $40~\mu\text{m}$ from the mill site. Specifically, even with dielectric encapsulation, regions directly subjected to direct ion bombardment exhibited significant degradation. Figure~\ref{fig:3} presents the radial Raman spectra of $\text{Al}_2\text{O}_3$-encapsulated $\text{MoS}_2$. Analysis of the spectra at $\text{P}4$ (the far-field reference point) reveals a slight narrowing of the $\text{E}'-\text{A}_1'$ peak gap compared to pristine, uncapped $\text{MoS}_2$, which we attribute to intrinsic strain induced by the $\text{Al}_2\text{O}_3$ capping layer itself. More critically, moving toward the patterned edge, the Raman spectra show a significant further broadening and shift of the characteristic modes, with the largest gap observed at $\text{P}1$ ($\mathbf{25~\text{cm}^{-1}}$). This demonstrates that the dielectric layer does not prevent the non-local transfer of strain and disorder, as substantial damage is observed extending approximately $\mathbf{40~\mu\text{m}}$ from the patterned site despite the encapsulation. While TMDC flakes located at a significant distance from the patterned area retained their optical properties post-annealing, the regions on the patterned microdisk were substantially affected, thereby preventing effective coupling to the microdisk. This indicated that direct ion bombardment, even through dielectric encapsulation, induced persistent defects that quenched the intrinsic optical response of the TMDC, hindering the formation of active photonic structures. This finding establishes the critical challenge of using conventional FIB to pattern high-quality photonic structures. This outcome suggests the need for alternative, non-damaging patterning techniques or refined ion beam parameters that selectively induce modifications without compromising the structural integrity or optical performance of the underlying 2D material. The dielectric encapsulation, while effective for preserving intrinsic material properties against environmental degradation, did not sufficiently shield the 2D TMDC from direct heavy ion impact, suggesting a need for materials with higher stopping power or novel shielding geometries for effective protection during FIB processing.

\begin{figure}[ht]
    \centering
    \includegraphics[width=1\linewidth]{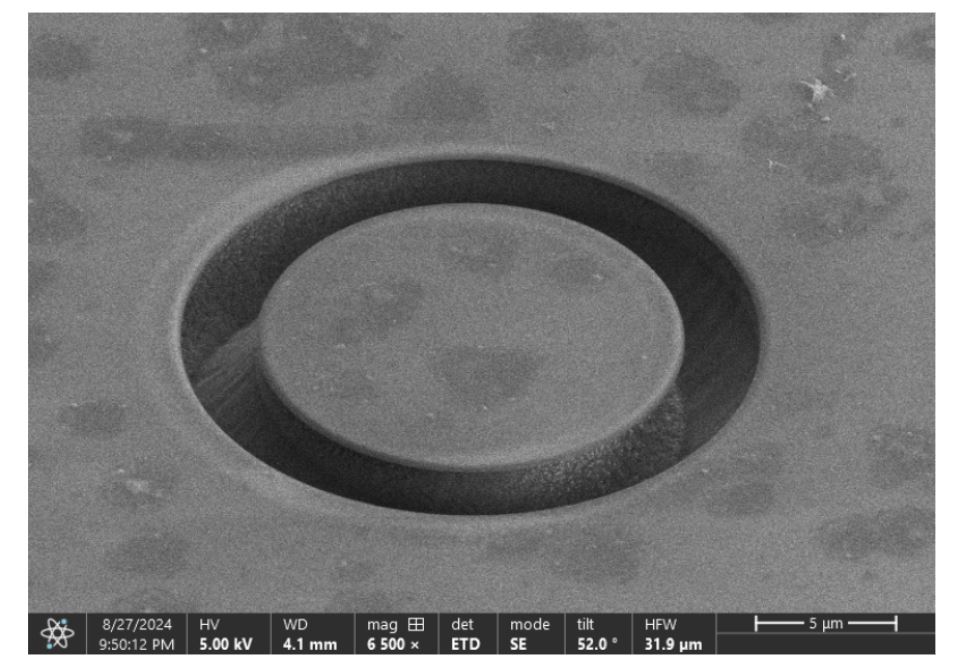}
    \caption{SEM image of a PMMA-capped 2D $\text{MoS}_2$ micro-resonator: Patterned using $\text{XeF}_2$-assisted Ga-FIB, the structure demonstrates precise material removal and ultra-smooth sidewalls critical for achieving low-loss coupling and high-quality factor (Q) in integrated photonic devices}
    \label{fig:5}
\end{figure}

In contrast to dielectric encapsulation, polymeric encapsulation with PMMA yielded promising results for retaining the optical properties of the 2D materials, as evident from our observations. Figure \ref{fig:4} illustrates our proposed fabrication approach, detailing the protective structure (a) and the specialized patterning method (b). This improvement is attributed to PMMA's ability to act as a sacrificial layer that absorbs the impact of gallium ions, thereby minimizing direct damage to the 2D TMDC and preventing substantial alteration of its optical characteristics. This protective effect of PMMA against ion-induced damage has been previously observed in other contexts involving delicate nanostructures \cite{nitipon_puttaraksa_cf1a23ea}, \cite{marcel_weinhold_029080e2}. The pliable nature of PMMA allows for energy dissipation from impinging ions, acting as a kinetic buffer that reduces the momentum transfer to the underlying 2D material layers. The PMMA layer effectively mitigates the deep penetration and scattering of incident ions, localizing the damage within the sacrificial polymer film and preserving the structural and electronic integrity of the encapsulated TMDC. This differential protection mechanism highlights the importance of selecting appropriate encapsulation materials based on the specific processing techniques and the desired outcome for integrated photonic devices. The enhanced stability observed with PMMA encapsulation, particularly against the detrimental effects of ion beam processing, presents a promising avenue for the fabrication of integrated photonic devices using 2D TMDCs.

\begin{figure*}[ht]
    \centering
    \includegraphics[width=1\linewidth]{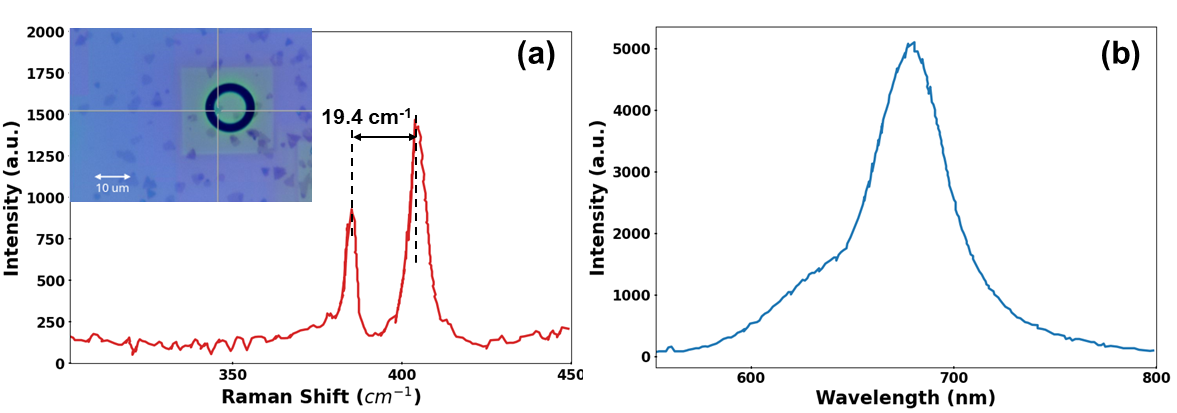}
    \caption{Preservation of intrinsic $\text{MoS}_2$ optical and structural quality via PMMA encapsulation: (a) Raman and (b) PL spectra measured adjacent to the patterned edge (inset) show full, unquenched exciton emission and characteristic Raman modes, establishing PMMA as a perfect protective layer against the influence of heavy ion}
    \label{fig:6}
\end{figure*}

To address the challenges of collateral damage and gallium ion implantation commonly associated with direct focused ion beam milling, $\mathbf{\text{XeF}_2\text{-assisted Ga ion beam direct patterning}}$ was employed to fabricate TMDC microdisks. The Gas Injection System (GIS) introduces the $\text{XeF}_2$ precursor gas directly into the chamber near the ion beam focus. The ion beam locally breaks the $\text{XeF}_2$ molecules, creating highly reactive fluorine radicals ($\text{F}^*$) which then chemically etch the exposed material. This combined physical sputtering and chemical etching process enhances the material removal rate while simultaneously reducing the reliance on high-energy physical sputtering alone, thereby minimizing collateral damage and especially Ga-ion implantation. This method demonstrated several significant advantages. The $\text{XeF}_2$-assisted process significantly reduced collateral damage and Ga ion implantation, which are prevalent issues with direct focused ion beam milling \cite{kyle_mahady_7cf2b611}, \cite{michael_g__stanford_86b43c7a}. The chemical etching component of this technique enabled anisotropic material removal with high precision, allowing for the definition of intricate microdisk geometries crucial for resonant optical modes \cite{michael_g__stanford_86b43c7a}.

The enhanced stability observed with $\text{PMMA}$ encapsulation, particularly against the detrimental effects of ion beam processing, combined with the $\text{XeF}_2$-assisted etching, presents a promising avenue for the fabrication of integrated photonic devices using 2D TMDCs. Figure \ref{fig:5} shows a high-resolution SEM image of a $\text{PMMA}$-capped $\text{MoS}_2$ micro-resonator patterned using this optimized technique. The image visually confirms the result of the anisotropic material removal: ultra-smooth sidewalls, which is critical for minimizing scattering losses and enhancing the quality factor of fabricated photonic resonators \cite{jangyup_son_b419e2dd}, \cite{jian_zhou_f55c1f27}.Furthermore, we provide definitive evidence of the protective capability of the $\text{PMMA}$ layer in Figure \ref{fig:6}, which presents the comparative Raman (a) and PL (b) spectra of the $\text{PMMA}$-encapsulated $\text{MoS}_2$ measured immediately adjacent to the patterned edge (inset). In stark contrast to the results shown in Figures \ref{fig:2} and \ref{fig:3}, the material retains its full, unquenched exciton emission and native Raman modes right up to the site of bombardment. This establishes $\text{PMMA}$ as a perfect protective layer against collateral $\text{FIB}$ damage.The combination of $\text{PMMA}$ encapsulation and $\text{XeF}_2$-assisted $\text{Ga}$ ion beam patterning provides a robust route for fabricating photonic devices using 2D materials. The polymer capping ensures the retention of the optical properties of the 2D materials, which is vital for subsequent integration into photonic devices where the intrinsic light-matter interaction of the TMDCs is critical for device functionality. Further, this approach simplified the fabrication process into a single step, combining patterning and etching, which offers advantages over complex multi-step processes conventionally employed \cite{fernando_j__urbanos_17a5dca2}. Lastly, the $\text{XeF}_2$-assisted focused ion beam etching method further minimizes the re-deposition of sputtered material, which often contaminates patterned structures and degrades device performance. For comparison, monolayer WSe$_2$ was synthesized via a similar CVD process using tungsten trioxide (WO$_3$) and selenium powders. Although the detailed results presented here focus on MoS$_2$, analogous measurements on WSe$_2$ revealed comparable responses under heavy ion exposure and PMMA encapsulation, confirming the generality of the observed effects across these 2D TMDCs. This refined patterning strategy thus enables the creation of high-quality integrated photonic components from 2D TMDCs, essential for advancing their application in quantum technologies and compact optical circuits.

\section{Conclusion}

This study systematically addressed the crucial challenge of preserving the intrinsic optical properties of 2D transition metal dichalcogenides during focused ion beam patterning, a fundamental step for their integration into next-generation photonic devices. We demonstrated that while FIB offers unparalleled precision for nanoscale fabrication and defect engineering, conventional dielectric encapsulation strategies proved inadequate in protecting these atomically thin materials from ion-induced damage. Regions directly exposed to gallium ion bombardment, even with dielectric capping, exhibited significant optical degradation, hindering effective device functionality.

In contrast, our findings highlight the superior protective capabilities of polymeric encapsulation using poly(methyl methacrylate). PMMA effectively acts as a sacrificial layer, absorbing the impact of gallium ions and mitigating damage to the underlying 2D TMDC, thereby preserving its critical optical characteristics. This protective mechanism, attributed to PMMA's pliable nature and high hydrogen content, ensures the retention of optical properties vital for device performance.

Furthermore, the adoption of XeF$_2$-assisted Ga ion beam direct patterning emerged as a pivotal advancement. This technique significantly reduces collateral damage and ion implantation, common drawbacks of conventional FIB milling. It enables precise anisotropic material removal, yielding ultra-smooth sidewalls essential for minimizing scattering losses and enhancing the quality factor of fabricated photonic resonators. This method simplifies the fabrication process, minimizes re-deposition, and facilitates the creation of intricate microdisk geometries crucial for resonant optical modes.

Collectively, the combination of PMMA encapsulation and XeF$_2$-assisted FIB patterning offers a robust, cost-effective, and scalable single-step fabrication route. This approach is instrumental in enabling the integration of 2D TMDCs into high-performance photonic devices, ensuring the maintenance of their intrinsic optical functionality. These advancements are critical for accelerating the development of novel applications in quantum technologies and compact optical circuits, paving the way for the realization of sophisticated 2D material-based integrated photonic systems.\\
\emph{Acknowledgement.} A.K. acknowledges funding from the Department of Science and Technology via the grants: SB/S2/RJN-110/2017, ECR/2018/001485 and DST/NM/NS-2018/49.

\bibliography{references.bib}

\end{document}